\documentclass[aps,eqsecnum,preprint,floats,epsf,epsfig,nofootinbib,letter]{revtex4}
\usepackage{epsfig}

\newlength{\defbaselineskip}
\newcommand{\setlinespacing}[1]%
           {\setlength{\baselineskip}{#1 \defbaselineskip}}

\def\DESepsf(#1 width #2){\epsfxsize=#2 \epsfbox{#1}}
\begin{document}

\begin{flushright}
\begin{tabular}{l}
 \\
CYCU-HEP-10-13
\end{tabular}
\end{flushright}
\vskip1cm

\title{ Modified Pad\'e Approach to the $S$-Wave Charmonium Spectroscopy in QCD}
\author{\sc Shu-Wei Chen}
\author{\sc Ching-Chang Lin}
\author{\sc Kwei-Chou Yang}

\affiliation{\vspace*{0.3cm} \normalsize\sl Department of Physics,
Chung Yuan Christian University, Chung-Li, Taiwan 320 \vspace*{2cm} }

\small
\begin{abstract}
\vspace{0.5cm}
We calculate the $S$-wave charmonium spectroscopy using the Hamiltonian with the non-relativistic QCD (NRQCD) potential. The logarithmic factor $\ln \mu r$, appearing in the next-to-leading order QCD loop corrections to the potential, is expanded about $r=1/\mu$, where $\mu$ corresponds to the typical charmonium scale. The resulting potential characterized by the Coulombic and linear components is consistent with the form of the Cornell potential. We obtain $\chi^2$ fitting results for the masses of the $S$-wave charmonium states, $\eta_c(1^1S_0)$, $J/\psi(1^3S_1)$, $\eta_c(2^1S_0)$, and $\psi(2^3S_1)$ in remarkable accordance with data. Our results successfully account for the hyperfine splitting for the $1S$ state as well as for the $2S$ state. We further use the three best fit parameters: the charm quark mass $m_c$, coupling constant $\alpha_s$ and the corresponding scale $\mu$ to predict the $S$-wave mass spectrum with $n\leq 6$. The hints for results are discussed.
\end{abstract}

\maketitle



\section{Introduction}

The heavy quarkonium (like charmonium, bottomonium, etc.) is a system where we can study the low-energy QCD in a systematic way. The heavy quarkonium satisfies the following hierarchy scales
$$ M\gg p\sim 1/r \sim Mv  \gg E \sim Mv^2,$$
where $M$ is the heavy-quark mass, $p$ the momentum transfer, $E$ the binding energy, $r$ the typical distance between the quarks, and $v$ the typical heavy quark velocity. It can be estimated that $v/c\sim 0.6$ for charmonium and $1/3$ for bottomonium \cite{Bodwin:1994jh,Braaten:1996ix}. Taking into account the above properties, one can introduce the nonrelativistic effective field theory which realizes a factorization at the Lagrangian between the high energy effects and low energy contributions. By integrating out the hard parts we can obtain the non-relativistic QCD (NRQCD) which is expanded in $p/M$ and $E/M$ \cite{Caswell:1985ui,Thacker:1990bm,Bodwin:1994jh}. Taking into account the fact that for the charmonia and bottomonia the typical scale $\mu\sim p$ associated to the inverse size of the system $1/r$ is satisfied by the relation $\mu (\sim 1/r) \gg E\sim {\rm \Lambda_{QCD}}$, the NRQCD can be further expanded in $Er$ and leads to an effective field theory which is the so-called potential non-relativistic QCD (pNRQCD) \cite{Pineda:1997bj,Brambilla:1999xf,Vairo:2000ia}.

Since the QCD has been widely accepted as a fundamental theory for the strong interactions, we have no doubt about that it should describe the spectroscopy of the heavy quarkonium. Nevertheless, in practice, it is still not so successful. On the other hand, the phenomenological quark models which mimic the QCD features seem to offer better results for the heavy quarkonium systems. The form of Cornell potential used in the phenomenological quark model \cite{Eichten:1974af,Eichten:1978tg,Eichten:1979ms}, where the potential is made of the Coulombic and linear parts, has been confirmed to be valid by the lattice calculation \cite{Bali:1999ai,Bali:2000zv}.

The $\bar{c} c$ charmonium states are usually denoted by the symbol $n^{2s+1}L_j$ with $n$ and $s$ being the principal and total spin quantum numbers, respectively. The $\eta_c(2^1S_0)$ was first measured in $B$ decays by Belle in 2002. Most potential model calculations predicted a much low value for its mass compared with the data. On the other hand, so far the hyperfine splitting for the $2S$ state ($m_{\psi(2^3S_1)}-m_{\eta_c(2^1S_0)}=49\pm 4$~MeV) as well as for the $1S$ state ($m_{J/\psi(1^3S_1)}-m_{\eta_c(1^1S_0)}=116.6\pm 1.2$~MeV) cannot be simultaneously calculated well compared with the data.  Although, the phenomenologically potential models \cite{Eichten:1978tg,Eichten:1979ms,Ebert:2002pp,Zeng:1994vj,Barnes:2005pb,Li:2009zu} can offer quite intuitive picture for $\bar{c} c$ charmonium spectroscopy, the deviation between the theoretical calculations and data is quite large, so that we cannot make further predictions accordingly. Even the old $\psi(4415)$, which was conventionally assigned as the $\psi(4^3S_1)$ quantum state, has been argued that it should be $\psi(5^3S_1)$ \cite{Li:2009zu}!

Recently, a number of interesting charmonium-like states, that are above the $D\bar{D}$ open-charm mass thresholds and named collectively as "$X,Y,Z$" mesons, have been found (see the discussions, {\it e.g.}, in Refs. \cite{Olsen:2008qw,Godfrey:2008nc}). So far, most of them are at odds with expectations of the $c\bar{c}$ states which had been predicted by conventional charmonium models. One may then suggest that some of the "$X,Y,Z$" particles are exotic. Existence of exotic states such as glueballs, hybrid mesons (the bound states of $\bar c c g$), molecules, and tetraquark mesons (the bound states of $\bar c c \bar c  c $), which go beyond the description of the naive
quark model, can offer the direct evidence concerning the confinement property of QCD. Nevertheless, so far, none of the exotic states can be well established. On the other hand, it is still difficult to assign any new observable in the conventional $\bar{c}c$ charmonium mass spectrum with the definite quantum state; many suggestions can be found in the literature. For instance, it was suggested that the $X(3940)$ may be the singlet state $\eta_c(3^1S_0)$. However the corresponding triplet state $3^3S_1$ is $\psi(4040)$ with mass $4039\pm1$ MeV, so that the assignment for $X(3940)$ implies a larger singlet-triplet mass splitting ($\simeq 100$ MeV) for radial number $n=3$ than that ($\simeq 50$ MeV) for $n=2$, which is one of the problems.

Motivated by the above reasons, in the present study we will try to obtain an evaluation for the $S$-wave charmonium spectroscopy starting with the pNRQCD Hamiltonian, instead of the phenomenologically potential model. We hope to clarify some ambiguities between observables and theoretical calculations. We use the QCD potential $V_{\rm QCD}(r)$, which can be obtained from matching NRQCD to pNRQCD \cite{Vairo:2000ia}. To solve the mass spectrum, we expand $\ln \mu r$, resulting from the QCD loop corrections to the potential, about $r=1/\mu$, where $\mu$
corresponds to the typical charmonium scale of order $m_c v$. The benefit of the expansion is that our resulting potential exhibits the form of the Cornell potential which was confirmed by the lattice calculation \cite{Bali:2000zv}. In our study, we have three parameters, the charm quark mass, $\alpha_s$ and $\mu$, which are related to the determination of charmonium masses. Using the modified Pad\'e approximation \cite{cnleungqu,cnleung} which is a numerical technique, we perform the best $\chi^2$-fit for the current data of masses of the well-established $S$-wave charmonium states, $\eta_c(1^1S_0)$, $J/\psi(1^3S_1)$, $\eta_c(2^1S_0)$, and $\psi(2^3S_1)$, and then use the fitted parameters, $m_c, \alpha_s$, and $\mu$ to further predict the full $S$-wave mass spectrum.   $\psi(2^3S_1)$ is usually denoted as the state $\psi(3683)$. As the fact that the total angular momentum $J$ is a conservation quantum number, the orbital angular momentum $L$ is actually not a good quantum number for the $\bar{c} c$ charmonium states. Therefore $\psi(3683)$ and $\psi(3777)$ could be the mixtures of $2^3S_1$ and $1^3D_1$ states \cite{Voloshin:2007dx}. Our result shows that the minimum $\chi^2$ is consistent with zero, which hints that the $S-D$ mixing may be negligible.

The Pad\'{e} approximation is to approximate a function $f(x)$, which is expanded in a Taylor series up to order $k$, by the ratio of two polynomials, one of order $M$ in the numerator, and another of order $N$ in the denominator, with $M+N=k$ \cite{pade:book1,pade:book2}. This ratio is called the Pad\'{e} approximant of $f(x)$. The technique of the Pad\'{e} approximation has the following advantages. First, it can accelerate the convergence of the usual Taylor expansion for a given function. Second, even for $x$ going beyond the radius of convergence of the Taylor's series of a given function $f(x)$, its Pad\'{e} approximant could well approximate the original function, i.e., physically it can be applied to the non-perturbative region. This method thus has been exploited in statistical physics, hadron phenomenology, quantum field theory \cite{ZinnJustin:1971ug,Queralt:2010sv,SanzCillero:2010mp,Masjuan:2009wy,Falkowski:2006uy,Peris:2006ds}, and recently in finding the solutions of general relativity \cite{Mroue:2008fu}.

The Pad\'{e} interpolation method, which is called the modified Pad\'{e} approximation here, was first proposed in Refs.~\cite{cnleungqu,cnleung} to explore physics in the non-perturbative region. In this modified approach, a single Pad\'{e} approximant is obtained by interpolating the weak and strong behaviors. We adopt this approach to study the charmonium spectroscopy. The QCD Hamiltonian is redefined as $\bar H(\beta)=H_C + \beta H_L$, where $H_C$ involves the Coulomb-like potential and $H_L$ contains the linear potential. We introduce the parameter $b$ to separate the kinetic energy term into two parts and then lump into $H_C$ and $H_L$ separately. As $\beta =1$, we have $\bar H(1)=H$, the physical Hamiltonian. We consider two limits, $\beta \gg 1$ and $\beta\ll 1$, to perform the perturbation calculation. After obtaining the results in the two limits, we can then get the physical eigenenergies corresponding to physical Hamiltonian $H$ using the Pad\'{e} interpolation. (See Sec. \ref{subsec:mpade} for the details.) In performing the fit, we also put the constraint on $b$, so that the numerical error in the approach due to the choice of $b$ is small enough ($\lesssim 2\%$). In general, when the radial number $n\leq 6$, the error is less than 7\% for $0.1<b<0.6$. The detailed discussion for numerical errors will be presented in Sec \ref{sec:num}.

The remaining of this paper is organized as follows. Together with an example, we will give a brief introduction to the methods of the conventional and modified Pad\'{e} approximations in Sec. \ref{sec:pade}. We formulate the modified Pad\'e approximant for charmonium masses in Sec. \ref{sec:formulation}. In Sec. \ref{sec:num}, the prediction for the $S$-wave charmonium mass spectrum, together with the best fit parameters, $m_c$, $\alpha_s$, and $\mu$, are given by minimizing $\chi^2$ fit. Sec. \ref{sec:conclusion} is our summary.

\section{The Pad\'{e} Approximation}\label{sec:pade}

\subsection{The conventional Pad\'{e} approximation}

The Pad\'{e} approximant  $f[M/N](x)$ of degree $(M,N)$, developed by H. Pad\'{e}, is an approximation of a given function $f(x)$ as a ratio of two power series:
\begin{eqnarray}\label{eq:pade-def}
f[M/N](x)&=&\frac{P_M(x)}{Q_N(x)}\\
 &=&\frac{p_0+p_1 x +p_2 x^2+p_3x^3+\cdots+p_M x^M}
 {1+q_1 x +q_2 x^2+q_3x^3+\cdots+p_N x^N}~,
\end{eqnarray}
where $P_M(x)$ and $Q_N(x)$ are polynomials of degrees $M$ and $N$, respectively. Assume that $f(x)$ is analytic around $x=0$ and has the Taylor expansion (or called the Maclaurin expansion) form:
\begin{eqnarray}\label{eq:taylor-1}
 f(x)=\sum_{i=0}^\infty a_i x^i.
\end{eqnarray}
Setting $f^{(n)}(0)=f^{(n)}[M/N](0)$ with $n=0,1,\dots, M+N$, one has
\begin{eqnarray}\label{eq:relation-1}
 \left( \sum_{i=0}^\infty a_i x^i\right) \left( \sum_{i=0}^M q_j x^j\right)
 - \left( \sum_{i=0}^N p_j x^j\right) ={\cal O} (x^{N+M+1}),
\end{eqnarray}
which can lead to $N+M+1$ linear equations:
\begin{eqnarray}\label{eq:relation-2}
 a_0 -p_0 =0 \nonumber\\
q_1 a_0 + a_1 -p_1=0  \nonumber\\
\vdots \nonumber \\
q_M a_{N-M} + q_{M-1} a_{N-M+1} + \cdots + a_N -p_N=0
\end{eqnarray}
and
\begin{eqnarray}
 q_M a_{N-M+1} + q_{M-1} a_{N-M+2} + \cdots + q_1 a_N+ a_{N+1}=0 \nonumber\\
 q_M a_{N-M+2} + q_{M-1} a_{N-M+3} + \cdots + q_1 a_{N+1} + a_{N+2}=0 \nonumber\\
\vdots \nonumber\\
 q_M a_{N} + q_{M-1} a_{N+1} + \cdots + q_1 a_{N+M-1} + a_{N+M}=0
\end{eqnarray}
From  the above independent equations, the $N+M+1$ coefficients, $p_i$ and $q_i$, can thus be determined.

For a given analytic function, its Pad\'{e} approximant of degree $(M,N)$ often gives much better approximation than truncating its Taylor series of degree $M+N$, and, moreover, the former may still work when the latter does not converge. Physically, this implies that not only the perturbative results can be further improved, but also it becomes possible to obtain a good estimate for nonperturbative phenomenologies.

\subsection{The modified Pad\'{e} approximation}

\begin{figure}[t]
\begin{center}
\epsfig{file=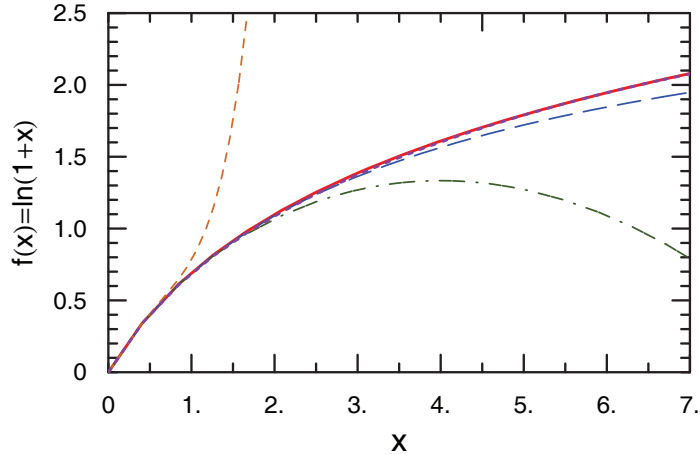,height=6cm,clip}
\end{center}
\caption{The graphs of the exact $f(x)=\ln(1+x)$ (solid curve), its Pad\'{e} approximants, $f[2/2](x)$ (long-dashed curve) and $f[3/1](x)$ (dot-dashed curve), modified Pad\'{e} approximant $f^{(2,1)}[3/1](x)$ (dotted curve), and Taylor's polynomial of degree 5 (short-dashed curve).}\label{fig1}
\end{figure}

In a practical calculation, we may not know well the full Taylor expansion of a given physical quantity at the specific point, {\it e.g.}, $x=0$, but just have its series up to a typical order. Following the idea by Leung and Murakowski \cite{cnleungqu}, the Pad\'{e} approximant of the function can be further improved if we know the truncated Taylor series of this function at the other analytic point. Here we would like to define the {\it modified} Pad\'{e} approximant for a given function as follows. For an analytic function $f(x)$ in the considered range of variable $x$, if we know its truncated Taylor series of degrees $r$ and $s$ respectively at $x=0$ and $x=x_0\not =0$,
\begin{eqnarray}\label{eq:taylor-2}
 f_{\rm Taylor}(x)= \sum_{i=0}^r a_i x^i + {\cal O}(x^{r+1}), \nonumber\\
 f_{\rm Taylor}(x)=\sum_{i=0}^s b_i (x-x_0)^i + {\cal O}(x^{s+1}),
\end{eqnarray}
in analogy to the relation given in Eq.~(\ref{eq:relation-1}), we can obtain $r+s+2 (=M+N+1)$ independent equations to determine the {\it modified} Pad\'{e} approximant $f^{(r,s)}[M/N](x)$:
\begin{eqnarray}\label{eq:pade-example}
f^{(r,s)}[M/N](x)=\frac{p_0+p_1 x +p_2 x^2+p_3x^3+\cdots+p_M x^M}
 {1+q_1 x +q_2 x^2+q_3x^3+\cdots+p_{N} x^{N}}~,
\end{eqnarray}
which may provide an accurate estimation for the original function in the entire range between the two expanding points.
Here we take the function $f(x)=\ln (x+1)$ as an example to illustrate the points. Expanding about the origin, which is equivalent to modeling the physically perturbation, the Taylor series of this function reads
\begin{eqnarray}\label{eq:ex-taylor-1}
f_{\rm Taylor}(x)= \sum_{i=1}^\infty (-1)^{i-1} \frac{x^i}{i},
\end{eqnarray}
which converges only for $-1<x \leq 1$. We can thus get the conventional Pad\'{e} approximants,
\begin{eqnarray}
 f[2/2](x)&=&\frac{\frac{x^2}{2} + x}{\frac{x^2}{6} + x + 1}, \nonumber\\
 f[3/1](x)&=&\frac{-\frac{x^3}{24} + \frac{x^2}{4} + x}{\frac{3 x}{4} + 1}.
\end{eqnarray}
On the other hand, we perform the Taylor expansion for the function at a large value of $x$, {\it e.g.} $x=6$, which is equivalent to the case of modeling the extremely strong coupling, reads
\begin{eqnarray}\label{eq:ex-taylor-2}
 f_{\rm Taylor}(x)=\log (7) + \frac{x -
        6}{7} - \frac{1}{98} (x -
          6)^2 + \frac{(x - 6)^3}{1029} - \frac{(x - 6)^4}{9604} +
  {\cal O}\left((x - 6)^5\right).
\end{eqnarray}
From the Taylor series results given in Eqs. (\ref{eq:ex-taylor-1}) and (\ref{eq:ex-taylor-2}), we obtain the modified Pad\'{e} approximants:
\begin{eqnarray}
 f^{(1,1)}[2/1](x)&=&\frac{\frac{1}{252}
 \left(6 - 7 \log (7) +
 \frac{12 (24 - 7 \log (7))}{-6 + 7 \log (7)}\right) x^2 + x}{\frac{(24 - 7 \log (7)) x}{3 (-6 + 7 \log (7))} + 1},\\
 f^{(2,1)}[3/1](x)&=&
 \frac{\frac{\left(132 - 7 \log (7) -
 \frac{54 (-12 - 7 \log (7))}{-24 + 7 \log (7)}\right) x^3}{3024} + \frac{1}{4} \
\left(-2 - \frac{12}{-24 + 7 \log (7)} -
 \frac{7 \log (7)}{-24 + 7 \log (7)}\right) x^2 + x}
 {\frac{(-12 - 7 \log (7)) x}{4 (-24 + 7 \log (7))} + 1}.\nonumber
\end{eqnarray}
In Fig. \ref{fig1} we plot the exact function $f(x)$, $f[2/2](x)$, $f[3/1](x)$, and $f^{(2,1)}[3/1](x)$, together with the result of truncated $f_{\rm Taylor}(x)$ expanding at $x=0$ up to $\mathcal{O}(x^5)$. More detailed numerical results are listed in Table~\ref{tab:pade}. The Taylor polynomial (corresponding to the perturbation) is valid only for $-1<x\leq 1$. The results of the Pad\'{e} approximations can offer reliable estimates extending to $x> 1$, corresponding to the non-perturbative region, and the accuracy can be enhanced if one increases the degree(s), $M$ and/or $N$, of the Pad\'{e} approximant. The modified Pad\'{e} approximant $f^{(2,1)}[3/1](x)$, which interpolates the results of the two different expanding points, differs from the exact value by no more than 1\% error between $x=0$ and $6$. Even for $f^{(1,1)}[2/1](x)$, the error is still less than 3\%. Extending to $f^{(2,2)}[3/2](x)$, the error becomes less than 0.1\%.

\begin{table}[t]
\caption{Comparison of the exact solution $f(x)=\ln(1+x)$ and its Pad\'{e} approximants as well as the Taylor polynomial, expanding at $x=0$ and up to ${\cal O}(x^5)$, where the modified Pad\'{e} approximants corresponding to the Taylor expansions both at $x=0$ and 6.} \label{tab:pade}
\addtolength{\arraycolsep}{0.1cm}
\begin{tabular}{|c|c|c|c|c|c|c|}
\hline $x$ & $f(x)$ (Exact)  & ~$f[2/2](x)$~ &  ~$f[3/1](x)$~ &  $f^{(1,1)}[2/1](x)$ &
$f^{(2,1)}[3/1](x)$ & $f_{\rm Taylor}$  to ${\cal O}(x^5)$
\\
\hline
0.5 & 0.4055 & 0.4054  & 0.4053  & 0.4146  & 0.4040 & 0.4073
\\
1   & 0.6931 &  0.6923 & 0.6905  & 0.7116  & 0.6878 & 0.7833
\\
2   & 1.0986 &  1.0909 & 1.0667  & 1.1209  & 1.0872 & $5.07$
\\
3   & 1.3863 & 1.3636  & 1.2692  & 1.4021  & 1.3751 & $35.85$
\\
4   & 1.6094 &  1.5652 & 1.3333  & 1.6172  & 1.6026 & $158.13$
\\
5   & 1.7918 & 1.7213  & 1.2719  & 1.7938  & 1.7896 & $502.92$
\\
6   & 1.9459 & 1.8462  & 1.0909  & 1.9459  & 1.9459 & $1291.2$
\\
\hline
\end{tabular}
\end{table}

\section{Formulations of heavy quarkonium masses}\label{sec:formulation}
\subsection{The Hamiltonian for $c\bar{c}$  bound states}
The Hamiltonian for the $c\bar{c}$ system expanding both in $\alpha_s$ and in $1/m_c$, determined from the perturbative QCD, is \cite{Vairo:2000ia,Pantaleone:1985uf}
\begin{eqnarray}
H&=&2m_c+H^{(0)}+V_{rel}
\end{eqnarray}
where $m_c$ is the mass of the charm quark at the scale $\mu$. $H^{(0)}$ is
\begin{eqnarray}
H^{(0)} = \frac{\vec{P}^2}{m_c}
+V_S,
\end{eqnarray}
including the kinetic energy and static potential up to ${\cal O}(\alpha_s^2)$
\begin{eqnarray}
V_S=-\frac{C_F\tilde{\alpha}_s}{r}-
\frac{C_F\beta_0\alpha_s^2}{2\pi}\frac{\ln r\mu}{r}~,
\end{eqnarray}
where $\vec{P}$ is the momentum of the charm quark, $\alpha_s$ the strong coupling constant, and
\begin{eqnarray}
\tilde{\alpha}_s &=& \left[1+\frac{a_1+\gamma_E\beta_0/2}{\pi}\alpha_s\right]\alpha_s,\nonumber\\
 C_F &=& \frac{4}{3},  ~~ \beta_0=11-\frac{2n_f}{3}\,,
\end{eqnarray}
with
\begin{eqnarray}
a_1=1.75,~~n_f=3~~\hbox{\rm for the charmonium \cite{y1,fjy9910399,nlo}},
\end{eqnarray}
and $\gamma_E\simeq 0.577216$ being the Euler constant. We consider the spin-dependently and spin-independently relativistic corrections $V_{rel}$ up to ${\cal O}(\alpha_s/m_c^2)$ and ${\cal O}(1/m_c^3)$, respectively,
\begin{eqnarray}
V_{rel}=V_{LS}+V_{T}+V_{hf} + V_{rel, K}\,.
\end{eqnarray}
Here $V_{LS}$, $V_T$ and $V_{hf}$ are the spin-orbit, tensor and
hyperfine (i.e. spin-spin) interactions, respectively, and $V_{rel,K}$ is the relativistically kinetic correction,  which are given by
\begin{eqnarray}
V_{LS} &=& \frac{3C_F\alpha_s}{2m_c^2r^3}\vec{L}\cdot\vec{S}\,, \\
V_T    &=& \frac{C_F\alpha_s}{4m_c^2 r^3}S_{12}\,, ~~~~{\rm with}~~
S_{12} = 3\frac{(\vec{r}\cdot\vec{S_1})(\vec{S_2}\cdot\vec{r})}{r^2}-\vec{S_1}\cdot \vec{S_2}\,, \\
V_{hf} &=& \frac{8\pi C_F\alpha_s}{3m_c^2}\vec{S_1}\cdot\vec{S_2}\delta^3(\vec{r})\,, \\
V_{rel,K} &=& -\frac{1}{4} \frac{P^4}{m_c^3} \,.
\end{eqnarray}

To solve the charmonium spectroscopy, we approximate the Taylor expansion of $\ln
r$ at $r=1/\mu$,
\begin{eqnarray}
\frac{\ln \mu
r}{r}&\approx&\frac{1}{r}\left\{ \left(r-\frac{1}{\mu} \right)\mu -\frac{1}{2} \left(r-\frac{1}{\mu} \right)^2\mu^2 +{\cal O} \left[\left(r-\frac{1}{\mu} \right)^3\mu^3\right] \right\}\\
&\approx&\frac{1}{r} \left(-\frac{3}{2}+2r\mu-\frac{1}{2}r^2\mu^2 \right),
\end{eqnarray}
with truncated series of degree 2.
For the charmonium, $\mu$ is the typical charmonium scale of order $1/r \sim m_c v$, where $v$ is the velocity of the charm quark. Consequently, we have
\begin{eqnarray}
H& \simeq &2m_c +\frac{P^2}{m_c}
 -\frac{C_F\tilde{\alpha}_s}{r}-\frac{C_F\beta_0\alpha_s^2}{2\pi}
 \frac{(-\frac{3}{2}+2r\mu-\frac{1}{2}r^2\mu^2)}{r} +V_{LS}+V_{T}+V_{hf} \label{eq:H-approx-1}\\
 &=&2m_c' +\frac{P^2}{m_c}
-\frac{\alpha}{r}+\lambda r+V_{LS}+V_{T}+V_{hf} + V_{rel,K} \label{eq:H-approx-2}\\
&=&2m_c' +H^{(0)}+V_{LS}+V_{T}+V_{hf} + V_{rel, K}, \label{eq:H-approx-3}
\end{eqnarray}
where
\begin{eqnarray}
 H^{(0)} &=& \frac{P^2}{m_c}-\frac{\alpha}{r}+\lambda r\,, \label{eq:para-def-1} \\
 \lambda &=& \frac{C_F\beta_0\alpha_s^2\mu^2}{4\pi} \,,  \\
\alpha&=& C_F\tilde{\alpha}_s-\frac{3}{2}\frac{C_F\beta_0\alpha_s^2}{2\pi}\,, \\
 m'_c &=& m_c-\frac{C_F\beta_0\alpha_s^2 \mu}{2\pi} \,.
\end{eqnarray}
In the spherical coordinate, it is known that
\begin{eqnarray}
 P^2 &=& (-i)^2 \nabla^2 =- \left(\frac{1}{r}\frac{\partial^2}{\partial r^2} r +\frac{1}{r^2 \sin \theta}
 \frac{\partial }{\partial \theta} \sin\theta \frac{\partial}{\partial\theta}+\frac{1}{r^2\sin\theta}\frac{\partial^2}{\partial\phi^2} \right)\nonumber\\
     &=&-\frac{1}{r}\frac{\partial^2}{\partial r^2} r
        -\frac{1}{r^2}
       \left( \frac{1}{ \sin \theta}\frac{\partial }{\partial \theta} \sin\theta
 \frac{\partial}{\partial\theta}+\frac{1}{\sin\theta}\frac{\partial^2}{\partial\phi^2} \right) \label{eq:P2-1}\\
     &\equiv&P_r^2+\frac{L^2}{r}  \,. \nonumber
\end{eqnarray}
We substitute Eq.~(\ref{eq:P2-1}) into Eq.~(\ref{eq:para-def-1}), and obtain
\begin{eqnarray}
 H^{(0)} &=&-\frac{1}{m_c} \left[\frac{1}{r}\frac{\partial^2}{\partial r^2} r
            +\frac{1}{r^2}(\frac{1}{ \sin \theta}\frac{\partial }{\partial \theta}
   \sin\theta\frac{\partial}{\partial\theta}+\frac{1}{\sin\theta}\frac{\partial^2}{\partial\phi^2})\right]
 -\frac{\alpha}{r}+\lambda r\nonumber\\
         &\equiv &\frac{1}{m_c} \left(P_r^2+\frac{L^2}{r^2} \right)-\frac{\alpha}{r}+\lambda r \,.\label{eq:H0}
\end{eqnarray}
A quantum state, with specified angular momentum quantum numbers $l$ and $m$, is
satisfied by
\begin{eqnarray}
 L^2|l,m\rangle =l(l+1)|l,m\rangle \,,\label{eq:angular-1} \qquad
 \langle \hat{n}|l,m \rangle =Y^m_l(\theta,\phi) \,, \label{eq:angular-2}
\end{eqnarray}
where $Y^m_l(\theta,\phi)$ are spherical harmonics. Therefore, in the calculation one can simply replace
$L^2$ by its eigenvalue $l(l+1)$ in Eq.~(\ref{eq:H0}), so that we have
\begin{eqnarray}
 H^{(0)}& \equiv &\frac{1}{m_c} \left[P_r^2+\frac{l(l+1)}{r^2}\right]-\frac{\alpha}{r}+\lambda r
 \nonumber\\
        &=&-\frac{1}{m_c r}\frac{\partial^2}{\partial r^2} r
           +\frac{l(l+1)}{m_c r^2}-\frac{\alpha}{r}+\lambda r \,.
\end{eqnarray}
Finally, we express $\bar{H}=H-2m_c'$ and concentrate on solving its eigenenergies. To perform the Pad\'{e} approximation study, we decompose the
Hamiltonian $\bar{H}=H-2 m_c^\prime$ into two parts, $H_C$ and $H_L$:
\begin{eqnarray}
 \bar{H}=H-2 m_c^\prime = H_C + H_L, \label{eq:effectiveH}
 \end{eqnarray}
 where
\begin{eqnarray}
 H_C&=& H_C^{(0)} +V_{LS}+V_{T}+V_{hf} \\
    &=&\
    b \left[ -\frac{1}{m_c r}
 \frac{\partial^2}{\partial r^2}r+\frac{l(l+1)}{m_c r^2} \right]
 -\frac{\alpha}{r} +V_{LS}+V_{T}+V_{hf}\,, \label{eq:HC}\\
 H_L&=& H_L^{(0)} +V_{rel,K}  \\
 &=& (1-b)\left[ -\frac{1}{m_c r}\frac{\partial^2}{\partial r^2} r+\frac{l(l+1)}{m_c r^2} \right]
 +\lambda r +V_{rel,K}\,, \label{eq:HL}
\end{eqnarray}
with $0<b<1$.
Here $H_C$ is the Hamiltonian contains the Coulomb potential, while $H_L$ involves the linear potential.

\subsection{ Masses of charmonia obtained from the modified Pad\'e approximation}\label{subsec:mpade}

As the Hamiltonian exhibited in Eqs. (\ref{eq:H-approx-1}) and (\ref{eq:H-approx-2}), at large distance, $r\gtrsim 1$ fm, the strong interaction potential is expected to rise linearly, so that we cannot treat the linear potential as perturbative term to obtain the solutions.  In the present study, we have one numerical parameter $b$, and three physical parameters: the charm quark mass $m_c$, strong coupling constant $\alpha_s$, and the scale $\mu$. We have checked that for $n\leq 6$ the {\it intrinsic} numerical errors are less than 7\%  if using $0.1<b<0.6$. The {\it intrinsic} error measures the difference between the exact value and its Pad\'e approximant. Putting the constraint on $b$, so that the intrinsic numerical errors are less than 2\% for all states with $n\leq 6$, we use the method of the modified Pad\'e approximation to approximate the masses of charmonia and then fit them with the current mass data for the $S$-wave charmonium states, $\eta_c(1^1S_0)$, $J/\psi(1^3S_1)$, $\eta_c(2^1S_0)$, and $\psi(2^3S_1)$ to determine the remaining three physical parameters. After that we can determined the $S$-wave mass spectrum.

To obtain the eigenenergies of $\bar{H}$,  we first define
\begin{eqnarray}
\bar{H}(\beta)=H_C + \beta H_L\,,\label{eq:effectiveH-2}
\end{eqnarray}
where $\beta$ is  a  real positive number. We can perturbatively solve the eigenenergies of $\bar{H}(\beta)$ in the two limits $\beta\ll1$ and $\beta\gg1$. As $\beta=1$, the eigenenergies of $\bar{H}(\beta=1)$ correspond to the real $c\bar{c}$ system. Once we have the results corresponding to $\beta\ll 1$ and $\beta\gg 1$, we can interpolate the two limits to obtain the eigenenergies of the real $c\bar{c}$ bound states by using the method of the modified Pad\'e approximation.  Note that $V_{rel,K}$ is contained in $H_L$ and its contribution is well under control in the perturbative calculation in the limit $\beta\ll 1$. In the calculation of the large $\beta$ limit, $V_{rel,K}$ is perturbatively small compared to $H_L^{(0)}$. If $V_{rel,K}$ is involved in $H_C$, its perturbatively correction to $H_C^{(0)}$ will be ${\cal O}(\alpha^2/(4b^3))$ and out of control for $b\simeq 0.275$ which is obtained in the later study. 
For the $S$-wave charmonium system, the mass spectrum is described by the eigenenergies $2m'_c+E_{n00s}^{(1,1)}[2/1](1)$, where $E_{n00s}^{(1,1)}[2,1](1)$ are the modified Pad\'e solutions of $\bar{H}(\beta=1)$ which will be explained below.

For a small $\beta$, using the Rayleigh-Schr\"odinger perturbation theory as given in the quantum mechanics textbooks \cite{jjs}, we obtain, for the $S$-wave states,
\begin{eqnarray}\label{eq:small-b}
E_{n00s}^< (\beta) &=& E^{<(0)}_{n00s} +E_{n00s}^{<(1)}\beta + \dots\,,
\end{eqnarray}
where $E^{<(0)}_{n00s} \equiv E_{n00s}$ is given by Eq.~(\ref{app:eigenenergies-r}) for which the detailed calculation can be found in Appendix~\ref{app:a}, and
\begin{eqnarray}\label{eq:small-b.1}
 E_{n00s}^{<(1)}& =& \langle \psi^C_{n00s}|H_L|\psi^C_{n00s} \rangle \nonumber\\
 &=&\int\psi_{n00s}^*(\vec{r}) H_L\psi_{n00s}(\vec{r}) r^2 \sin\theta dr d\theta d\phi\nonumber\\
&=&\int^\infty _0
\left(-\frac{1-b}{m_c} \bar{R}_{ns}(r)\frac{1}{r}\frac{\partial^2}{\partial r^2} r
\bar{R}_{ns}(r)r^2 + \lambda \bar{R}_{ns}(r) r \bar{R}_{ns}(r) r^2 \right)dr\,.
\end{eqnarray}
Here $\bar{R}_{ns}(r)$ is defined by Eq. (\ref{app:WF}), and $|\psi^{C}_{nlms}\rangle$ are the eigenkets of the Hamiltonian $H_C$, given by
\begin{eqnarray}
 |\psi^{C}_{nlms}\rangle &=&  |\psi_{nlms}\rangle\otimes |s,s_z;s_1,s_2\rangle \,,
\end{eqnarray}
where $|\psi_{nlms}\rangle$ and $|s,s_z;s_1,s_2\rangle$ respectively correspond to the spatial and spin parts of the wave functions and  we have
\begin{eqnarray}
  \langle \vec{r}|\psi_{n00s}\rangle = \psi_{n00s}(\vec{r})\,,\label{eq:WF-space}
\end{eqnarray}
(see also Eq.~(\ref{app:WF}) for the detailed expression).

For a large $\beta$, we can rewrite the Hamiltonian in the following form
\begin{eqnarray}\label{eq:large-b}
 \bar{H}(\beta) = \beta \left[ H^{(0)}_L + V_{rel,K} +\frac{1}{\beta}(H^{(0)}_C+V_{LS}+V_T+V_{hf}) \right]\,.
 \end{eqnarray}
For the $S$-wave states, the energy spectrum $E^>_{n00s}$ corresponding to the large $\beta$ limit can be expanded in power series with respect to $1/\beta$:
 \begin{eqnarray}
  E^>_{n00s}(\beta) &=&\beta \left(  E_{n00s}^{>(0)}+ \widetilde{E}_{n00s}^{>(1)}+ E_{n00s}^{>(1)}\frac{1}{\beta}
    + \dots\right)\,.
  \end{eqnarray}
Here $ E_{n00s}^{>(0)}$ is the zeroth corrections, and $ \widetilde{E}^{>(1)}_{n00s}$ and $ E^{>(1)}_{n00s}$ are the first corrections. They can be evaluated by the perturbation theory and are
\begin{eqnarray}
 E_{n00s}^{>(0)} &=& -\left[ \frac{(1-b)\lambda^2}{m_c} \right]^{\frac{1}{3}} r_n\,, \\
 \widetilde{E}_{n00s}^{>(1)} &=&
 \langle \psi^{L(0)}_{n}|V_{rel,K}|\psi^{L(0)}_{n} \rangle \nonumber\\
 &=&\int\psi^{L(0)}_{n}(r) \, V_{rel,K}\, \psi^{L(0)}_{n}(r) r^2 \sin\theta drd\theta d\phi\nonumber\\
 &=&\int^\infty _0
 \Bigg\{B_n \frac{Ai\Big((\frac{m_c}{1-b} \lambda)^{\frac{1}{3}}r +r_n \Big)}{r}\frac{-1}{4 m_c^3 r}\frac{\partial^4}{\partial r^4}
 \left[r B_n \frac{Ai\Big((\frac{m_c}{1-b} \lambda)^{\frac{1}{3}}r +r_n \Big)}{r} \right]
 \Bigg\} 4\pi r^2 dr , \\
 E_{n00s}^{>(1)} &=&
 \langle \psi^{L(0)}_{n}|H_C|\psi^{L(0)}_{n} \rangle \nonumber\\
 &=&\int\psi^{L(0)}_{n}(r) (H^{(0)}_C+V_{hf})\psi^{L(0)}_{n}(r) r^2 \sin\theta drd\theta d\phi\nonumber\\
 &=&\int^\infty _0
 \Bigg\{B_n \frac{Ai\Big((\frac{m_c}{1-b} \lambda)^{\frac{1}{3}}r +r_n \Big)}{r}\frac{-b}{m_c r}\frac{\partial^2}{\partial r^2}
 \left[r B_n \frac{Ai\Big((\frac{m_c}{1-b} \lambda)^{\frac{1}{3}}r +r_n \Big)}{r} \right] \nonumber\\
&& +
\left[B_n  \frac{Ai\Big((\frac{m_c}{1-b} \lambda)^{\frac{1}{3}}r +r_n \Big)}{r} \right]^2
\left[- \frac{\alpha}{r} +\frac{4\pi C_F \alpha_s}{3m_c^2}\left(s(s+1)-\frac{3}{2}\right)
\frac{\delta(r)}{4\pi r^2}  \right]
 \Bigg\} 4\pi r^2 dr , ~~~~~~
\end{eqnarray}
where the contributions due to $V_{LS}$ and $V_T$ vanish (see Appendix~\ref{app:a}), $E_{n00s}^{>(0)}$ are the eigenenergies of $H_L$ with $l=0$, and the corresponding eigenfunctions are
\begin{eqnarray}
\psi^{L(0)}_{n} (r)&=& B_n \frac{Ai\Big((\frac{m_c}{1-b}\lambda)^{\frac{1}{3}}r +r_n \Big) }{r}\,,
\end{eqnarray}
with $A_i$ being the so-called Airy function, the normalization:
\begin{eqnarray}
B_n &=& \left( \int^\infty _{r_n}
\frac{[Ai(r)]^2}{(\frac{m_c}{1-b}\lambda)^{\frac{1}{3}}}4\pi dr\right)^{-1/2}\,,
\end{eqnarray}
and $r_n$ the roots of the Airy function.

Now we compute the energy spectrum for $S$-wave $c\bar{c}$ bound states using the modified Pad\'{e}
approximation. The eigenenergies of the Hamiltonian $\bar{H}(\beta)$ are approximately by the modified Pad\'e
approximants:
 \begin{eqnarray}
E^{(1,1)}_{n00s}[2/1](\beta)&=&\frac{p_0 + p_1 \beta +p_2 \beta^2}{1+q_1\beta}~.\label{eq:pade-cc-def}
  \end{eqnarray}
The coefficients $p_0,p_1,p_2$, and $q_1$ can be determined in the following way. Comparing with Eq.~(\ref{eq:small-b}), for a small $\beta$, we have
\begin{eqnarray}\label{eq:pade_expansion-1}
E_{n00s}^{(1,1)}[2/1](\beta)
               &=&p_0 -(p_1 -p_0 q_1)\beta+ {\cal O}(\beta^2) \\
               &=&E_{n00s}^{<(0)}+E_{n00s}^{<(1)}\beta + {\cal O}(\beta^2) \,,
\end{eqnarray}
On the other hand, comparing with Eq.~(\ref{eq:large-b}), for a large $\beta$, we get
\begin{eqnarray}\label{eq:pade_expansion-2}
E_{n00s}^{(1,1)}[2/1](\beta)
&=&\frac{p_2}{q_1}\beta+\left(-\frac{p_2}{q_1^2}
+\frac{p_1}{q_1}\right) + {\cal O}\left(\frac{1}{\beta} \right)\\
&=& E_{n00s}^{>(0)}\beta+ E_{n00s}^{>(1)}
    + {\cal O}\left(\frac{1}{\beta} \right) \,.
\end{eqnarray}
We therefore obtain the relations:
\begin{eqnarray}
 p_0=E_{n00s}^{<(0)}\,, \\
 p_1-p_0 q_1=E_{n00s}^{<(1)}\,, \\
 \frac{p_2}{q_1}= E_{n00s}^{>(0)}\,, \\
 -\frac{p_2}{q_1^2} +\frac{p_1}{q_1}= E_{n00s}^{>(1)}\,.
\end{eqnarray}
and arrive at the eigenenergies of real $S$-wave $c\bar{c}$ bound states:
\begin{eqnarray}
E_{n00s}^{(1,1)}[2/1](\beta=1)&=&\frac{p_0 +p_1 +p_2}{1+q_1}\,,
\end{eqnarray}
with
\begin{eqnarray}
p_0 &=&E_{n00s}^{<(0)}\,, \\
p_1 &=&\frac{E^{<(0)}_{n00s}  E^{>(0)}_{n00s}-E^{<(1)}_{n00s} E^{>(1)}_{n00s}}
  {E^{<(0)}_{n00s}-E^{>(1)}_{n00s}}\,,\\
p_2 &=&\frac{ E^{>(0)}_{n00s}(-E^{<(1)}_{n00s}+  E^{>(0)}_{n00s})}{E^{<(0)}_{n00s}- E^{>(1)}_{n00s}}~,\\
q_1&=&\frac{-E^{<(1)}_{n00s}+ E^{>(0)}_{n00s}}{E^{<(0)}_{n00s}-  E^{>(1)}_{n00s}}\,.
\end{eqnarray}

\section{Numerical analysis and discussions}\label{sec:num}

We have only three parameters, $m_c$, $\alpha_s$, and $\mu$ to be related to the physical masses, and one numerical parameter $b$. The $b$ is related to the {\it intrinsic} error for the Pad\'e approximant compared to its true value. In the fit, we put the constraint on $b$, so that the {\it intrinsic} error of the modified Pad\'e approach is small enough $\lesssim 2\%$ for the states with the radial quantum number $n \leq 6$. In general, the error is less than 7\% for $0.1<b<0.6$.  (We will further discuss the {\it intrinsic} error later.) We adopt the masses of four well-measured $S$-wave charmonium states, $\eta_c(1^1S_0)$, $J/\psi(1^3S_1)$, $\eta_c(2^1S_0)$, and $\psi(2^3S_1)$ \cite{pdg}, as inputs to determine $m_c$, $\alpha_s$, and $\mu$. In Particle Data Group (PDG) \cite{pdg},  $\psi(2^3S_1)$ is denoted by $\psi(2S)$ or $\psi(3683)$, which will be discussed later.

Under the {\it intrinsic} error $\lesssim 2\%$, we perform the best $\chi^2$ fit which is defined by minimizing
\begin{eqnarray}
\chi^2 = \chi^2_{10} + \chi^2_{11} + \chi^2_{20} + \chi^2_{21}\,,
\end{eqnarray}
with
\begin{eqnarray}
\chi^2_{ns} = \left(\frac{m_{n00s}^{expt}-m_{n00s}^{th}}{\delta m_{n00s}^{expt}} \right)^2\,,
\end{eqnarray}
where $m_{n00s}^{expt}\pm\delta m_{n00s}^{expt}$ are the experimental charmonium masses and $m_{n00s}^{th}=2m'_c+E^{(1,1)}_{n00s}[2/1](1)$  the theoretical predictions that we have calculated in this paper. We find that the minimum of $\chi^2$ is $\chi^2_{min}=0.26$ as well as  $b\simeq 0.275$ corresponding to the almost smallest intrinsic error in the fit.  Our results, in good agreement with the data, can successfully account for the hyperfine splitting for the $1S$ state as well as for the $2S$ state.
The best fit values for parameters are
\begin{eqnarray}
m_c (\mu)&=&1.517~{\rm GeV}\,, \label{eq:parameter-1}\\
\alpha_s(\mu)&=&0.273\,, \label{eq:parameter-2}\\
 \mu&=&1.843~{\rm GeV} \label{eq:parameter-3}\,.
\end{eqnarray}
One that note that $m_c, \alpha_s$, and $\mu$ are not really physical parameters since an isolated charm quark cannot be observed. The renormalization scale $\mu$, which is adopted to separate the potential into several parts, is chosen to be the quantity that charmonium becomes stable, so that after some combination we have $m_c \to m_c'$ and $\alpha_s \to \tilde\alpha_s$ as shown in Eqs. (3.12) and (3.13).

It is interesting to note that the Coulombic and linear potentials defined in Eq.~(\ref{eq:H-approx-2}) are then obtained to be
\begin{eqnarray}
-\frac{\alpha}{r}+\lambda r = -\frac{0.288}{r} + (0.241 \, {\rm GeV}^2) \ r, \label{eq:cornell-like}
\end{eqnarray}
while in the Cornell potential model the potential is parametrized as $V=-a/r + e r$, with $a\simeq 0.52$ and $e\simeq 0.18$ GeV$^2$ \cite{Eichten:1979ms}. For the Cornell potential, $a$ is usually identified by $C_F \alpha_s$. Nevertheless, our $\alpha$ is
$$\alpha = C_F \left[1+\frac{a_1+\gamma_E\beta_0/2}{\pi}\alpha_s\right]\alpha_s -\frac{3}{2}\frac{C_F\beta_0\alpha_s^2}{2\pi} .$$
Using the obtained parameters, we can further get the masses of higher $S$-wave states. The results are given in Table \ref{tab:mass}. For comparison, we also list the current data assignments \cite{pdg} and some other theoretical results \cite{Chen:2000ej,Liao:2002rj,Brambilla:2001fw,Ebert:2002pp,Zeng:1994vj}. It was known that $\psi(3683)$ and  $\psi(3770)$ could be the mixtures of $2^3S_1$ and $1^3D_1$ states due to the fact that, instead of $L$, the total angular momentum $J$ is a conserved quantum number; $L$ can be broken by some relativistic effects, for which especially the operator of the tensor force does not commute with $L^2$ for states with $S=1$ (see the discussions in Ref. \cite{Voloshin:2007dx}). However, we see that the fitted $\chi^2_{\rm min}$ is consistent with zero, which may hint that the $S$-$D$ mixing effect is negligible. One should note that the relatively large $e^+ e^-$ width of the $\psi(3770)$ is difficult to understand if it is a pure $1^3D_1\ \bar{c}c$ state. This problem can be solved if $\psi(3770)$ has an admixture of $2^3S_1\ \bar{c}c$ state \cite{Barnes:2005pb}.
Our results show that if the smallness of $S$-$D$ mixing effects can be applied for higher radial excited states, the singlet-triplet splitting mass difference is about 50 MeV for $n\geq 3$.
Conventionally, the observables $\psi(4040)$ and $\psi(4160)$ were assigned as $3^3S_1$ and $2^3D_1$ states, respectively. However, our calculation suggests that the $\psi(4160)$ may be dominated by the $3^3S_1$ state. It has been noted that the $\psi(4160)$ has a much larger $e^+ e^-$ width so that it may have a significant $S$-wave $\bar{c}c$ component \cite{Barnes:2004cz}. In the flux-tube model, the light hybrid charmonium states lie $\sim$ 4.1 GeV and it was suggested that $\psi(4040)$ and $\psi(4160)$ may be the strong mixtures of the hybrid charmonium and $\psi(3S)$ \cite{Close:1995eu}, which can explain why $\Gamma^{e^+ e^-}(\psi(4040))\simeq \Gamma^{e^+ e^-}(\psi(4160))$.  If so, the $X(4160)$ might be further assigned as the $\eta_c(3^1S_1)$. The $X(4160)$ with a mass of $(4156\pm 29)$ MeV$/c^2$ and a total width of $\Gamma=(139^{+113}_{-65})$ MeV$/c^2$ was seen by Belle in the $D \bar{D}^*$ recoiling from the $J/\psi$ in the annihilation process $e^+ e^-\to J/\psi D^{*}\bar{D}^{*}$ \cite{Abe:2007sya}. We obtain the mass for the $4^3S_1$ to be $4475.1\pm 13.5$ MeV, which is  about 50 MeV larger than $\psi(4415)$ which is conventionally assigned as the $4^3S_1$ state. These discrepancies can be further clarified by including higher order corrections in the calculation.

\begin{table}[ht]
\caption{ The predictions for the $S$-wave charmonium spectroscopy (in units of MeV), compared with the results from the current data assignments (PDG) and from theoretical calculations by lattice QCD (Lattice), perturbative QCD-based (PQCD), and phenomenological quark model (QM). The theoretical errors are estimated in Table \ref{tab:pade-error}.}\label{tab:mass}\vspace{0.3cm}
\begin{tabular}{|c|l|c|c|c|c|c|}
\hline State ($n ^{2s+1}L_j$) & PDG\cite{pdg} & This work& Lattice\cite{Chen:2000ej,Liao:2002rj}& PQCD\cite{Brambilla:2001fw} & QM\cite{Ebert:2002pp}
& QM\cite{Zeng:1994vj}\\
\hline $\eta_c$($1^1S_0$)& $2980.5 \pm 1.2$   &$2980.5\pm0.0$  &$3014\pm4$&3056&2979&3000\\
\hline $J/\psi$($1^3S_1$)& $3096.916\pm 0.011$&$3096.9\pm0.0$ &$3084\pm4$ &3097&3096&3100\\
\hline $\eta_c$($2^1S_0$)& $3637\pm 4$        &$3634.9\pm3.0$    &~~$3707\pm20$&--- &3583&3670\\
\hline $\psi$($2^3S_1$)  & $3686.093\pm 0.034$ &$3686.1\pm3.0$&~~$3780\pm43$&---  &3686&3730\\
\hline $\eta_c$($3^1S_0$)&      ---           &$4068.4\pm8.0$          &---     &---& 3991&4130\\
\hline $\psi$($3^3S_1$)  &$4039\pm1\footnotemark[1] $     &$4118.6\pm8.0$  &---  &--- & 4088 &4180\\
\hline $\eta_c$($4^1S_0$)&      ---           &$4424.5\pm13.5$          &---    &--- &---&---\\
\hline $\psi$($4^3S_1$)  &$4421\pm4\footnotemark[2]$     &$4475.1\pm13.5$ &---    &---&---&4560\\
\hline $\eta_c$($5^1S_0$)&    ---             &$4730.0\pm18.8$          &---    &---  &---  &---\\
\hline  $\psi$($5^3S_1$) &     ---            &$4781.8\pm18.8$          &---     &---  &---  &---\\
\hline $\eta_c$($6^1S_0$)&    ---             &$5000.4\pm23.9$          &---     &--- &---  &---\\
\hline  $\psi$($6^3S_1$) &     ---            &$5052.0\pm23.9$          &---     &---  &---  &---\\
\hline
\end{tabular}
\footnotetext[1]{It is called $\psi(4040)$.}
\footnotetext[2]{It is called $\psi(4415)$.}
\end{table}
\vskip1cm

For estimating the numerical uncertainties for our predictions \footnote{Without existence of the spin-spin (hyperfine) interaction, one could solve numerically the Schroedinger equation for the "spatial part" of the Hamiltonian $H$, i.e., determine its spatial wave functions. However, it is highly nontrivial if there exists the spin-spin interaction and one would like to fit numerical solutions to the data. The spin-spin interaction is relevant to explain the mass splitting between
$\eta_c(1 ^1S_0)$ and  $J/\psi(1 ^3S_1)$ and between  $\eta_c(2 ^1S_0)$ and  $J/\psi(2 ^3S_1)$ , which cannot be computed well in the literature so far. }, we take into account the eigen-solutions for the spatial Hamiltonian $H^{(0)}$ with $l=0$, i.e., the full Hamiltonian $H$ without the spin-spin interaction term and with $l=0$ :
\begin{eqnarray}
 \left( -\frac{1}{m_c r}\frac{\partial^2}{\partial r^2} r
      -\frac{\alpha}{r}+\lambda r \right) \psi_{n}^{(0)} = E_{n}^{(0)} \psi_{n}^{(0)}\,.
\end{eqnarray}
As in Eqs. (\ref{eq:HC}) and (\ref{eq:HL}), we introduce the parameter $b$ and then split $H^{(0)}$ into two parts, $H_C^{(0)}$ and $H_L$. Substituting the values for $m_c$, $\lambda$, $\alpha$, given in Eqs. (\ref{eq:parameter-1}), (\ref{eq:parameter-3}), (\ref{eq:cornell-like}) and adopting $b=0.275$, we numerically solve the above equation. Table \ref{tab:pade-error} compares the modified Pad\'e results with numerically exact eigenenergies. We see that the modified Pad\'e approach yields approximations which are not larger than 1.5\% of the exact solutions for states with $n\leq 6$.

\begin{table}[ht]
\caption{Error estimate in the modified Pad\'e approach. The eigenenergies are in units of MeV.}\label{tab:pade-error}\vspace{0.1cm}
\begin{center}
\begin{tabular}{|c|r|r|r|c|}
\hline { State ($n S$)} & $E_{n}^{(0)}$ (Exact) & $E_{n}^{(0)}$ (Pad\'e) & Pad\'e-Exact (error)
& (Pad\'e-Exact)/Exact\\
\hline $1S$ & $603.26$  & $603.29$   & $0.03$ & 0.01\%\\
\hline $2S$ & $1253.87$ & $1256.86$  &2.99    & 0.2\% \\
\hline $3S$ & $1761.25$ & $1769.20$  & $7.95$& 0.5\% \\
\hline $4S$ & $2202.15$ & $2215.61$  & $13.46$& 0.6\% \\
\hline $5S$ & $2601.77$ & $2620.58$  & $18.81$& 0.7\% \\
\hline $6S$ & $2972.23$ & $2996.09$  & $23.86$& 0.8\% \\
\hline
\end{tabular}
\end{center}
\end{table}

\section{Summary}\label{sec:conclusion}

The $S$-wave charmonium spectroscopy has been calculated by considering the Hamiltonian with the non-relativistic QCD potential.  For the next-to-leading order QCD loop corrections to the potential, we expand the logarithmic factor $\ln \mu r$ about $r=1/\mu$, where $\mu$ corresponds to the typical charmonium scale of order $m_c v$, so that the QCD potential can be modeled as the Coulomb plus linear form, which is consistent with the Cornell potential.  In our approach, we have performed the best $\chi^2$ fit by comparing the current mass data of the $S$-wave charmonium states, $\eta_c(1^1S_0)$, $J/\psi(1^3S_1)$, $\eta_c(2^1S_0)$, and $\psi(2^3S_1)$, with their modified Pad\'e approximants. Our results, in good agreement with the data, can successfully account for the hyperfine splitting for the $1S$ state as well as for the $2S$ state.
The fitted parameters are $m_c (\mu)= 1.517~{\rm GeV},
\alpha_s(\mu) =0.273$, and $\mu =1.818~{\rm GeV}$, consistent well with the ranges that one usually used. Using then these three parameters we have further predicted the $S$-wave mass spectrum with $n\leq 6$.

\vskip 2.5cm

\acknowledgments
We are grateful to C.H. Chen and C.W. Kao for useful comments. We also would
like to thank N. Brambilla for correspondence.
This research was supported in part by the National Center for Theoretical Sciences and the
National Science Council of R.O.C. under Grant No. NSC96-2112-M-033-004-MY3 and No. NSC99-2112-M-003-005-MY3.

\appendix

\section{Eigenenergies and eigenfunctions of the Hamiltonian $H_C$} \label{app:a}

The eigenenergies and corresponding eigenfunctions of the Hamiltonian operator $H_C$ can be calculated by using the Rayleigh-Schr\"odinger perturbation theory. We decompose $H_C$ into two Hermitian parts, $H_C^{(0)}$ and the rest,
\begin{equation}
         H_C=H^{(0)}_C+ (V_{LC}+V_T+V_{hf})\,,
\end{equation}
where
\begin{eqnarray}
V_{LS}&=&\frac{3C_F\alpha_s}{2m_c^2r^3}\vec{L}\cdot\vec{S}\,,\\
V_T &=& \frac{ C_F\alpha_s}{4 m_{c}^2 r^3} S_{12}
 =\frac{C_F\alpha_s}{4 m_{c}^2 r^3}
 \left( 3\frac{(\vec{r}\cdot\vec{S_1})(\vec{S_2}\cdot\vec{r})}{r^2}-\vec{S_1}\cdot \vec{S_2} \right) \,,\\
V_{hf} &=& \frac{8\pi C_F \alpha_s}{3m_c ^2} \vec{S_1}\cdot \vec{S_2} \delta^3(\vec{r})
 \equiv \frac{4\pi C_F \alpha_s}{3m_c ^2}\left[s(s+1)-\frac{3}{2}\right]\delta^3(\vec{r})\,.
\end{eqnarray}
To solve the eigenvalue problem of $H_C$, we instead consider the following Hamiltonian function
\begin{equation}
         H_C(\kappa)=H^{(0)}_C+ \kappa (V_{LC}+V_T+V_{hf})\,,
\end{equation}
so that we have $H_C = H_C(1)$, where $\kappa$ is a continuous real parameter.

\subsection{The corrections due to the spin-obit and tensor interactions in $S$-wave states of Hamiltonian $H_C$}

It is obviously that $\langle V_{LS}\rangle=0$, where $\langle \cdots \rangle$ means the expectation value of the $S$-wave states, while $\langle V_T\rangle$ is given by
\begin{eqnarray}
\langle V_T\rangle &=& \frac{ C_F\alpha_s}{4 m_{c}^2 r^3}\langle S_{12}\rangle \nonumber\\
     &=&\frac{C_F\alpha_s}{4 m_{c}^2 r^3}\langle 3\frac{(\vec{r}\cdot\vec{S_1})(\vec{S_2}\cdot\vec{r})}{r^2}-\vec{S_1}\cdot \vec{S_2}\rangle \,,
\end{eqnarray}
where
\begin{eqnarray}
\lefteqn{\langle 3\frac{(\vec{r}\cdot\vec{S_1})(\vec{S_2}\cdot\vec{r})}{r^2}-\vec{S_1}\cdot\vec{S_2}\rangle }\nonumber\\
 &=&
 \langle \int d^3r |\psi_{n00s}(\vec{r})|^2\frac{3(\vec{r}\cdot\vec{S_1})(\vec{S_2}\cdot\vec{r})-\vec{S_1}\cdot \vec{S_2}r^2}{r^2} \rangle \nonumber\\
 &=& \langle \int d^3r |\psi_{n00s}(\vec{r})|^2\times \Biggl[
 \frac{3(S_{1x}~ x+S_{1y}~y+S_{1z}~ z)(S_{2x}~ x+S_{2y}~ y+S_{2z}~z)}{r^2}-\frac{\vec{S_1}\cdot\vec{S_2}~r^2}{r^2}\Biggr] \rangle \,, \label{a6}
\end{eqnarray}
with $\psi_{n00s}(\vec{r})$ being the spatial part of the wave function for the Hamiltonian $H_C$, given in Eq.~(\ref{app:WF}), which is spin-dependent but independent of $\phi$ since $m=0$.
In terms of the variables of spherical polar coordinates, $r$, $\theta$ and $\phi$:
\begin{eqnarray}
x&=&r \sin\theta \cos\phi\,, \\
y&=&r \sin\theta \sin\phi\,, \\
z&=&r \cos\theta\,,
\end{eqnarray}
we have
\begin{eqnarray}
\int xy d\Omega&=&\int r^2 \sin^3\theta \cos\phi \sin\phi
d\theta d\phi\nonumber\\
&=&0=\int yz d\Omega=\int xz d\Omega\,. \label{a1}\\
\int x^2 d\Omega&=&\int r^2 \sin^3\theta \cos^2\phi
d\theta d\phi\nonumber\\
&=&\frac{4\pi}{3}r^2=\int y^2 d\Omega=\int z^2 d\Omega\,. \label{a5}
\end{eqnarray}
Using the results of Eqs.~(\ref{a1}) and (\ref{a5}) to the angular integral in Eq.~(\ref{a6}), we obtain
\begin{eqnarray}
\int d\Omega (\vec{r}\cdot\vec{S_1})(\vec{S_2}\cdot\vec{r})&=&\int
d\Omega (S_{1x} x+S_{1y} y+S_{1z} z)(S_{2x} x+S_{2y} y+S_{2z} z)\nonumber\\
&=&\int d\Omega\frac{1}{3}r^2(S_{1x} S_{2x}+S_{1y} S_{2y}+S_{1z}
S_{2z})\nonumber\\
&=&\int d\Omega\frac{1}{3}r^2 \vec{S_1}\cdot\vec{S_2}\,,
\end{eqnarray}
and therefore
\begin{eqnarray}
\lefteqn{\langle V_T\rangle} && \nonumber\\
&=&
 \langle \frac{C_F\alpha_s}{4 m_{c}^2}\int d^3r |\psi_{n00s}|^2\frac{3(S_{1x} x+S_ {1y}y+S_{1z} z)(S_{2x}
x+S_{2y} y+S_{2z} z)-\vec{S_1}\cdot \vec{S_2}r^2}{r^5} \rangle \nonumber\\
&=&
 \langle \frac{C_F\alpha_s}{4 m_{c}^2}\int d^3r |\psi_{n00s}|^2\frac{\vec{S_1}\cdot \vec{S_2}r^2-\vec{S_1}\cdot
\vec{S_2}r^2}{r^5} \rangle =0\,.
\end{eqnarray}

\subsection{The corrections in $S$-wave states of Hamiltonian $H_C$}

Since the spin-orbit and tensor interactions can be neglected for the $S$-wave states, following the standard approach, the eigenenergies $E_{nlms}(\kappa)$ of $H_C(\kappa)$ can be determined in terms of the perturbation expansion:
\begin{eqnarray}
 E_{nlms}(\kappa) &=& E_{nlms}^{(0)}+E_{nlms}^{(1)}\kappa +E_{nlms}^{(2)}\kappa^2+E_{nlms}^{(3)}\kappa^3+\cdots~,\\
    E_{nlms}^{(0)}&=&-\frac{m_c \alpha^2}{4b}\frac{1}{n^2}~,\\
    E_{nlms}^{(1)}&=& \langle \psi_{nlms}^{C(0)}|
 \frac{4\pi C_F \alpha_s}{3m_c ^2}\left[s(s+1)-\frac{3}{2}\right]\delta^3(\vec{r})|\psi_{nlms}^{C(0)}
 \rangle \nonumber\\
  &=&\int\psi^{C(0)}_{nlm}(\vec{r})
   \frac{4\pi C_F \alpha_s}{3m_c ^2}\left[s(s+1)-\frac{3}{2}\right]\delta^3(\vec{r})\psi^{C(0)}_{nlm}(\vec{r})
   r^2 \sin\theta dr d\theta d\phi \\
 \Biggl( &=&\int^\infty _0 R^{(0)}_n(r)
 \frac{4\pi C_F \alpha_s}{3m_c ^2}\left[s(s+1)-\frac{3}{2}\right](2l+1)\frac{\delta(r)}{4\pi r^2} R^{(0)}_n(r)r^2 dr,  \hskip0.3cm {\rm for}\  m=0 \Biggr)\,, \nonumber\\
 E_{nlms}^{(2)}&=& \sum_{k=1,\dots\atop k\neq n ~~}\frac{V_{nklms}V_{knlms}}{E^{(0)}_{nlms}-E^{(0)}_{klms}}\,, \\
 E_{nlms}^{(3)}&=& \sum_{k=1,\dots\atop k\neq n ~~}
 \sum_{g=1,\dots\atop g\neq n ~~}
 \frac{V_{nklms}V_{kglms}V_{gnlms}}
 {(E^{(0)}_{nlms}-E^{(0)}_{klms})(E^{(0)}_{nlms}-E^{(0)}_{glms})}-\sum_{k=1,\dots\atop k\neq n ~~}
 \frac{V_{nnlms}V_{nklms}V_{knlms}}{(E^{(0)}_{nlms}-E^{(0)}_{klms})^{2}},
\end{eqnarray}
where $E_{nlms}^{(0)}$ and $|\psi^{C(0)}_{nlms}\rangle$ are respectively the eigenenergies and eigenkets of $H^{(0)}_C$, and
\begin{eqnarray}
 V_{nklms} &=& V_{knlms}^* \nonumber\\
       &=& \langle \psi_{nlms}^{C(0)}|\frac{8\pi C_F \alpha_s}{3m_c^2}\vec{S_1}\cdot \vec{S_2}\delta^3(\vec{r})|\psi_{klms}^{C(0)} \rangle \nonumber\\
   \Biggl( &=&
   \int^\infty _0 R^{(0)}_n(r)
 \frac{4\pi C_F \alpha_s}{3m_c ^2}\left[s(s+1)-\frac{3}{2}\right](2l+1)
 \frac{\delta(r)}{4\pi r^2} R^{(0)}_k(r)r^2 dr,  \hskip0.3cm {\rm for}\  m=0 \Biggr). ~~
\end{eqnarray}
The states $|\psi^{C(0)}_{nlms}\rangle$ read
\begin{eqnarray}
 |\psi^{C(0)}_{nlms}\rangle &=&  |\psi^{C(0)}_{nlm}\rangle\otimes |s,s_z;s_1,s_2\rangle \,,
\end{eqnarray}
where
\begin{eqnarray}
 \langle \vec{r}|\psi^{C(0)}_{nlm}\rangle &=& \psi^{C(0)}_{nlm} (\vec{r})
 = R^{(0)}_n(r)Y^m_l(\theta,\phi)\nonumber\\
 &=&Y^m_l(\theta,\phi) \left (\frac{m_c\alpha}{n b}\right)^{\frac{3}{2}}
 \sqrt{\frac{(n-l-1)!}{2n [(n+l)!]}}
 \left(\frac{m_c\alpha r}{n b} \right)^l e^{-\frac{\alpha m_c}{2 n b}r}L^{2l+1}_{n-l-1}
 \left(\frac{m_c\alpha r}{n b}\right) . ~~
\end{eqnarray}
In the calculation, we introduce the transformation, called the Pad\'e approximation, to accelerate the convergence of the perturbative series of $E_{nlms}(\kappa)$ which are approximately presented as rational functions, $E^{PA}_{nlms}[2/1](\kappa)$,
\begin{eqnarray}
 E_{nlms}(\kappa)\simeq E^{PA}_{nlms}[2/1](\kappa)&=&
 \frac{p^C_0+p^C_1\kappa+p^C_2\kappa^2}{1+q^C_1 \kappa},
\end{eqnarray}
where $p^C_{0,1,2}$ and $q^C_1$ can be determined by the values of $E_{nlms}^{(0)}$, $E_{nlms}^{(1)}$, $E_{nlms}^{(2)}$, and $E_{nlms}^{(3)}$.

Therefore, for the $S$-wave states, the eigenenergies of the Hamiltonian $H_C$ are approximately to be
\begin{eqnarray}\label{app:eigenenergies-r}
 E_{n00s} = E_{n00s}(1)\simeq
 E^{PA}_{n00s}[2/1](1) &=&
  \frac{p^C_0+p^C_1+p^C_2}{1+q^C_1 },
\end{eqnarray}
and the corresponding wave functions are
\begin{eqnarray}
 |\psi^{C}_{n00s}\rangle &=&  |\psi_{n00s}\rangle\otimes |s,s_z;s_1,s_2\rangle \,,
\end{eqnarray}
where up to the second order
\begin{eqnarray}
  \langle \vec{r}|\psi_{n00s}\rangle = \psi_{n00s}(\vec{r}) &\simeq& \psi^{C(0)}_{n00}(\vec{r}) + \psi^{C(1)}_{n00s}(\vec{r}) + \psi^{C(2)}_{n00s}(\vec{r})\nonumber\\
 &=& \sqrt{\frac{1}{4\pi}} \left[ R^{(0)}_n(r)+R^{(1)}_{ns}(r)+R^{(2)}_{ns}(r) \right]
 \equiv\sqrt{\frac{1}{4\pi}}\bar{R}_{ns}(r)\,, \label{app:WF}
\end{eqnarray}
with
\begin{eqnarray}
\psi^{C(0)}_{n00}(\vec{r}) &=&\sqrt{\frac{1}{4\pi}}(\frac{m\alpha}{n b})^{\frac{3}{2}}\sqrt
{\frac{(n+1)!}{2n (n!)}} \left(\frac{m\alpha r}{n b} \right)
e^{-\frac{\alpha m_c}{2 n b}r}L^{1}_{n-1} \left(\frac{m\alpha r}{n b} \right)~,\\
 \psi^{C(1)}_{n00s}(\vec{r}) &=&\sum_{k=1,\dots\atop k\neq n ~~}
 \frac{V_{nk00s}}{E^{(0)}_{n00s}-E^{(0)}_{k00s}}\psi^{C(0)}_{k00}(\vec{r})\nonumber \\
 &=&
 \sum_{k=1,\dots\atop k\neq n ~~}
 \frac{V_{nk00s}}{E^{(0)}_{n00s}-E^{(0)}_{k00s}}R^{(0)}_k(r)\sqrt{\frac{1}{4\pi}}
 =\sqrt{\frac{1}{4\pi}}R^{(1)}_{ns}(r)\,, \\
\psi^{C(2)}_{n00s}(\vec{r}) &=&\sum_{k=1,\dots\atop k\neq n ~~}
 \sum_{g=1,\dots\atop g\neq n ~~}
 \frac{V_{ng00s}V_{gk00s}}{(E^{(0)}_{n00s}-E^{(0)}_{k00s})(E^{(0)}_{n00s}-E^{(0)}_{g00s})}\psi^{C(0)}_{k00}(\vec{r})\nonumber\\
& &-\sum_{k=1,\dots\atop k\neq n ~~}
 \frac{V_{nk00s}V_{nn00s}}{(E^{(0)}_{n00s}-E^{(0)}_{k00s})^{2}}\psi^{C(0)}_{k00}(\vec{r}) \nonumber\\
 &=& \sqrt{\frac{1}{4\pi}}R^{(2)}_{ns}(r)\,.
\end{eqnarray}

\end{document}